\documentclass[aps,prb,superscriptaddress,showpacs,floatfix,twocolumn,amsmath]{
revtex4-1}
\usepackage{graphicx}
\usepackage{bm,amsmath,amssymb,mathrsfs,dcolumn}
\usepackage{subfigure}
\usepackage{units}
\usepackage{hyperref}
\usepackage[usenames]{color}
\hypersetup{pdfborder=0 0 0,colorlinks=true,citecolor=blue,linkcolor=blue}
\newcommand{\angstrom}{\mbox{\normalfont\AA}}

\begin{document}


\pacs{61.50.Ah,65.40.-b,88.40.-j}

\title{Room temperature dynamic correlation between methylammonium molecules in 
lead-iodine based perovskites: An \textit{ab-initio} molecular dynamics 
perspective}

\date{\today}

\author{Jonathan Lahnsteiner}
\author{Georg Kresse}
\affiliation{University of Vienna,
Faculty of Physics and Center for Computational Materials Science, Sensengasse 
8/12,
A-1090 Vienna, Austria}

\author{Abhinav Kumar}
\author{D.D. Sarma}
\affiliation{Solid State and Structural Chemistry Unit, Indian Institute of 
Science, 560012 Bangaluru, India}

\author{Cesare Franchini}
\author{Menno Bokdam}
\email{menno.bokdam@univie.ac.at}
\affiliation{University of Vienna,
Faculty of Physics and Center for Computational Materials Science, Sensengasse 
8/12,
A-1090 Vienna, Austria}

\begin{abstract}
The high efficiency of lead organo-metal-halide perovskite solar cells 
has raised many questions about the role of the methylammonium (MA) 
molecules in the Pb-I framework. Experiments indicate that the MA 
molecules are able to 'freely' spin around at room temperature even though they 
carry an intrinsic dipole moment. We have performed large supercell (2592 
atoms)  finite temperature ab-initio molecular dynamics calculations to study 
the correlation between the molecules in the framework.  An 
underlying long range 
anti-ferroelectric ordering 
of the molecular dipoles is observed. The dynamical correlation between 
neighboring molecules shows a maximum around room temperature in 
the mid-temperature phase. In this phase, the rotations are slow enough 
to (partially) couple to neighbors via the Pb-I cage. This results in a 
collective motion of neighboring molecules in which the cage acts as the 
mediator. At lower and higher temperatures the motions are less correlated.

\end{abstract}

\maketitle


\hyphenation{organo-metal}


The spectacular rise of perovskite photovoltaics\cite{NREL16} has sparked 
much research effort into the physical mechanisms behind these materials' good 
photovoltaic performance. From a solid state physics perspective, the answer 
seems simple: the well suited electronic structure of methylammonium 
lead-iodide (MAPbI$_3$). With a band gap of $\sim$1.6 
eV\cite{Hirasawa:jpsj94,Tanaka:ssc03,Papavassiliou:sm95,Lee:sc12,Stoumpos:ic13} 
and a high absorption coefficient\cite{Hirasawa:jpsj94,Lee:sc12,Stoumpos:ic13}, 
it is expected to be a high efficiency solar cell material for the solar 
radiation spectrum observed on earth.\cite{Shockley:jap61} 
Density Functional Theory (DFT) calculations have shown that an $s$-$p$ mixture 
in the valence band maximum 
(VBM) and $p$-states in the conduction band minimum (CBM), combined with a 
direct band gap lead to an absorption coefficient up to an order of magnitude 
higher than 
GaAs.\cite{Yin:am14} However, the crystal structure of MAPbI$_3$ is completely 
different from GaAs. It even possesses temperature depended dynamical 
contributions arising from the organic constituent. The perovskite structure 
of MAPbI$_3$ is composed of three iodine atoms (monovalent anions) combined 
with a lead atom (divalent cation) and a CH$_3$NH$_3$ (MA) molecule (monovalent 
cation). The Pb-I framework forms the perovskite structure out of PbI$_6$ 
octahedra and the molecules are trapped in the cavities. Whether the cubic 
perovskite structure 
is stable or 
becomes orthorhombic or tetragonal is determined by the 
temperature\cite{Stoumpos:ic13} combined with a balance between the size of the 
molecule and the Pb-I bond length.\cite{Li:acb08} NMR experiments have 
indicated that at high temperatures ($>$300~K) the MA molecule exhibits 
complete orientational disorder.\cite{Wasylishen:ssc85,Baikie:jmca:15} This 
means that the 
MA molecules have enough kinetic energy to overcome the rotational barriers and 
can rotate in their 'cage'. Around room temperature this process has a typical 
relaxation time of $\sim5$~ps.\cite{Chen:pccp15} These rotations apparently do 
not effect the charge 
carriers in the system, since very long electron-hole diffusion 
lengths have been reported.\cite{Xing:sc13,Stranks:sc13} From an electronic 
point 
of view this is 
not surprising, because the Pb-I framework is electronically decoupled from 
the molecule.\cite{Filippetti:prb14} The VBM/CBM are well separated from the 
HOMO/LUMO of the molecule, respectively. Therefore all conducting states for 
holes and electrons are dominated by the framework. At the same time the 
dispersion at the VBM and CBM is large, resulting in low effective 
masses.\cite{Umari:sr14} However, this picture applies to 0~K and does not take 
into account the 
dynamics of the MA molecules, their order or lack thereof at room temperature 
(RT). A partial alignment of these polar molecules could result in polar 
domains. Piezoresponse force microscopy measurements 
seem to indicate that polar domains form and that 
spontaneous polarization occurs at RT.\cite{Kutes:jpcl14,Kim:jpcl15} A DFT 
based 
analysis indicates that at high temperature ferroelectricity is suppressed by 
the large 
configurational entropy and only below 50~K would ferroelectric (FE) ordering 
be 
present.\cite{Filippetti:jpcl15} A strong argument against a polar 
structure is the absence of second harmonic generation in time-resolved 
pump-probe measurements.\cite{G:jpcl16} The 
presence of polarized domains at RT is however still debated.

In a recent work we showed that polar phonons in MAPbI$_3$ increase the 
static 
dielectric constant to $\sim30$, but they are too slow to effectively 
screen the excitonic state in the system.\cite{Bokdam:sr16}  
The rotational dynamics of 
the polar MA molecules and the corresponding screening occurs at 
even lower frequencies.\cite{Perez:jpcc15,Brivio:prb15,Mattoni:jpcl16} 
Therefore even when 
the temperature is raised to RT, \textit{i.e.} when the 
molecules rotate, the excitons are not screened by polar 
phonons nor by rotating polar molecules.\cite{Bokdam:sr16}  
However, understanding of the dynamics of the molecules at finite 
temperatures is necessary to describe thermal 
transport\cite{Hata:nanol16,Sheng:prb16} as well as the screening of free 
charge 
carriers.\cite{Neukirch:nanol16}  
The dynamics and (dis)order of the MA molecules in MAPbI$_3$ have been 
extensively investigated, both experimentally and theoretically. Experiments 
showed the ability of molecules to reorient at RT with relaxation times in the 
range (1-­‐10~ps).\cite{Wasylishen:ssc85,Poglitsch:jcp87,Onoda-Yamamuro:jpcs90,
Bakulin:jpcl15,Chen:pccp15,Leguy:natc15} The dynamics of the MA molecules in 
the PbI framework were also studied using \textit{ab-initio} molecular dynamics 
reporting relaxation times in roughly the same picosecond time 
scale.\cite{Mosconi:pccp14,Frost:aplm14,Carignano:jpcc15,Goehry:jpcc15,
Meloni:natc16,Even:nanos16} Most \textit{ab-initio} studies were performed with 
small systems. A recent study of the 
relaxation time as a function of temperature using a reasonably large super 
cell indicates an absence of spatial 
correlation at RT.\cite{Meloni:natc16} However, a precise assessment of finite 
size effects in \textit{ab-initio} calculations is still missing. MA dynamics 
in large supercells has been studied by classical molecular 
dynamics.\cite{Mattoni:jpcc15} In agreement with experiment\cite{Chen:pccp15} a 
transition from the anti-ferroelectric ordering in the orthorhombic phase 
to 
thermally activated directional disorder in the tetragonal and cubic phases was 
reported.\cite{Mattoni:jpcc15} However, the accuracy of model potentials is 
questionable and there is the need for an \textit{ab-initio} validation. In 
particular, no systematic analysis of the spatial and dynamical correlation has 
been reported.

In this work we 
study the long range order of the MA molecules in MAPbI$_3$ by means of large 
scale 
ab-initio molecular dynamics (MD) calculations at finite temperatures. We 
show that the dynamics of the molecules is not a completely random thermal 
motion. Besides the possible dipole-dipole interaction between the intrinsic 
dipole moments of the molecules\cite{Frost:nanol14}, the order of the 
molecules is also mediated by long range cage stress and strain effects as well 
as deformations 
or rotations of the PbI$_6$ octahedra and their long range order. In the low 
temperature 
($<$150~K) phase this leads to an anti-ferroelectric arrangement of the 
molecules.\cite{Swainson:jssc:03,Chi:jssc05,Baikie:jmca:13} 
At higher temperatures the ordering pattern 
changes. 

The paper is organized as follows. In the next section, the details of the 
computational method are presented. Hereafter, we present the different super 
cell structures and their 
implications. We discuss in section \ref{secB} the dynamics of 
the 
molecules individually and in section \ref{secC} the correlation with their 
neighbors. We discuss the results in \ref{secE}, and in the last 
section we summarize our findings in conclusions.


\section{Computational method}
For the first-principles molecular 
dynamics calculations we use a plane-wave basis and the 
projector augmented wave (PAW) 
method\cite{Blochl:prb94b} as implemented in the {\sc vasp} 
code\cite{Kresse:prb93,Kresse:prb96,Kresse:prb99}. The 
PBEsol (Perdew, Burke, 
Ernzerhof modified for solids)\cite{Perdew:prl08} functional is used. 
Relatively 
shallow pseudo-potentials are used, for 
Pb the 
($6s^26p^2$), for I the ($5s^25p^5$), for C the ($2s^22p^2$) and for N the 
($2s^22p^3$) orbitals are included in the valence. This makes it possible to 
set a relatively low energy cut-off of 250~eV for the 
plane-wave basis. 

Super cells of dimension $n\times{}n\times{}n$, $n=2,4,6$ were constructed out 
of 
pseudo-cubic unit cells (12 atoms) with experimental lattice constants 
($a,b=6.3115$~\angstrom, $c=6.3161$~\angstrom)\cite{Stoumpos:ic13} as 
described 
in Ref. 
\onlinecite{Bokdam:sr16}. In order to construct an unbiased starting 
structure, all molecules in the supercell are randomly (R) rotated over three 
axis before starting the MD run. In the supercell all atoms were allowed 
to move while keeping the lattice vectors fixed.

Gaussian smearing with 
$\sigma=0.05$~eV is used to broaden the one-electron levels. The Brillouin 
zone is sampled by the 
$\Gamma$ point only for the large $4\times4\times4$ (4-cell) and 
$6\times6\times6$ (6-cell)
supercells and by a  ($2\times2\times2$) $\Gamma$-centered 
Monkhorst-Pack grid for the $2\times2\times2$ supercell (2-cell). 
 Using only a single k-point for the 2-cell results in 
significant errors in the calculated forces. The 
Kohn-Sham orbitals 
are 
updated in the 
self-consistency cycle until an energy convergence of $10^{-4}$~eV is 
obtained. A Langevin thermostat\cite{Allen:book91} is applied 
to simulate a canonical ensemble (at constant temperature). The 
trajectory is formed by propagating 
the structure with the calculated Hellmann-Feynman forces with time 
steps of  2.0/2.0/3.0~fs and
increased hydrogen masses of  4/4/8~a.u. for the 2/4/6-cell, 
respectively. The 
hydrogen masses are increased to allow for a larger time step for the large 
cell-sizes. Additionally in the 6-cell the Pb and I masses have 
been decreased to 20~a.u. These changes in the masses are possible since the 
DFT functional does not 
depend on the 
mass and since the partition function factorizes into a momentum and position 
dependent part ($\rho(X,P)=e^{-{\rm U}(X)\beta}e^{-P^2/(2M)\beta}$) so that 
results for the configurational part $X$ are independent of the chosen masses 
($M$). The down side is that the dynamical properties (velocities, average 
reorientation 
times, etc.) are no 
longer 
exact. However, differences in the dynamical properties between calculations 
where the same masses are used remain meaningful. 

\begin{figure} [b!]
\includegraphics[width=8.cm]{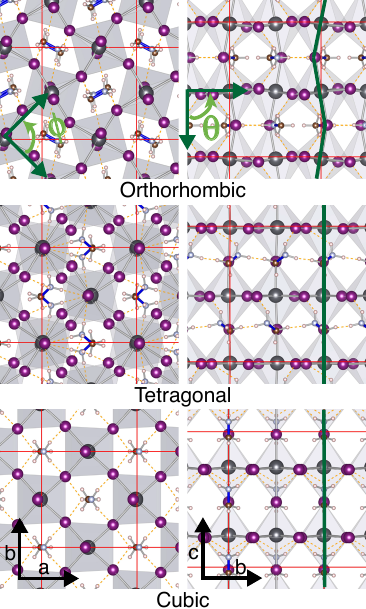}
\caption{(color online) The three different MAPbI$_3$ crystal phases. The unit 
cells (48 atoms, 4 molecules, solid red lines) are indicated, except for the 
cubic phase where a 
$\sqrt{2}\times\sqrt{2}\times2$ supercell of the pseudo-cubic unit cell is 
shown. The dashed orange lines show the hydrogen bonds between the NH$_3$ 
groups and the iodines. The left/right column shows a top/side view, 
respectively. The polar coordinates reference frame \{$\phi,\theta$\} to 
describe the orientation of the C-N bond of the MA molecule is drawn on the 
top.}
\label{fig1a}
\end{figure}

Unit cells for the different crystal phases have been constructed based on 
experimental crystal structures determined by X-ray diffraction. For the 
orthorhombic structure the Pb-I positions of 
Baike \textit{et al.}\cite{Baikie:jmca:13} were used 
($a=8.8362$~\angstrom, $b=8.5551$~\angstrom, $c=12.5804$~\angstrom) and for the 
tetragonal structure the Pb-I positions of Stoumpos \textit{et 
al.}\cite{Stoumpos:ic13} were 
used ($a=8.8490$~\angstrom, $b=8.8490$~\angstrom, $c=12.6420$~\angstrom). The 
molecules were placed in an arbitrary but unpolarized pattern and a high 
temperature MD (500~K, Pb-I framework kept fixed) was performed, 
allowing 
the molecules to reorient themselves. Several low 
energy structures were picked from the MD trajectory and for those all atoms 
were thoroughly relaxed (forces $<10^{-5}$ eV/$\angstrom$) into their 
instantaneous 
ground state while keeping the lattice vectors fixed. The resulting 
lowest energy configurations are shown in 
Figure \ref{fig1a}.

\section{Molecular Dynamics: unit and supercells}
\label{secA}

In the MAPbI$_3$ structure the different MA molecules can 
interact with each 
other, either directly via the electrostatic forces (they are cations 
and have an intrinsic 
dipole moment) or indirectly by their collective pushing and pulling on the 
Pb-I 
framework. The potential energy related to dipole-dipole 
interactions between the 
different molecules falls off like $r^{-3}$, and we expect that this 
interaction only 
influences the nearest neighbors. However, it is clear that a single unit 
cell, be it orthorhombic (4 formula units), tetragonal (4 formula units) or 
cubic (1 formula unit) is prone to finite size 
effects/self-interaction errors and might therefore be artificially stabilized 
in a particular ordering pattern. If these effects are ignored and one starts 
with 
the structures obtained by X-ray diffraction\cite{Baikie:jmca:13,Stoumpos:ic13} 
for 
the three different phases, well ordered patterns are obtained as shown in 
Figure \ref{fig1a}. These structures agree (at various levels of 
accuracy) with 
the orthorhombic\cite{Menendez:prb14,Filip:natc14,Filip:prb14} and 
tetragonal\cite{Quarti:com14,Amat:nanol14} molecular ordering patterns found in 
other first principles studies. The three crystal phases have distinct 
molecular ordering patterns:

The \textit{orthorhombic phase} has an antiferroelectric (AFE) ordering 
pattern of the 
MA molecules, with two molecules lying flat in the $ab$-plane with a 
$\sim60^{\circ}$ angle between them. The layer on top has the same relative 
orientation in the plane, but the molecules are $180^{\circ}$ degree 
rotated resulting in a zero net dipole moment. The PbI$_6$ octahedra are 
rotated by 
$\sim15^{\circ}$ degree in the plane, one clockwise, and the other 
counterclockwise. 
The octahedra in the layer below have the same rotations in the 
$ab$-plane. An alternating 
$\pm 11^{\circ}$ tilt from the c-axis of neighboring 
octahedra results in a zigzag pattern in the 
c-direction.

In the \textit{tetragonal phase} the octahedra are aligned to the c-axis, 
but in the $ab$-plane the octahedra in the two layers do not have 
the same rotations. They are rotated by $\sim12^{\circ}$ 
degree, again one clockwise and the other counterclockwise. But in the second 
layer the order is reversed, i.e., counterclockwise and clockwise. In 
this Pb-I framework the MA molecules have an orthogonal orientation (when 
projected) in the 
$ab$-plane to all their 
nearest neighbors and have a $\sim30^{\circ}$ out of the $ab$-plane tilt. The 
resulting angle between all nearest neighboring molecules thereby becomes 
$\sim75^{\circ}$. The total MA pattern here is slightly FE.

In the \textit{cubic phase} the octahedra are almost not rotated and 
all 
molecules are 
aligned in a FE ordering pattern. At the corresponding temperature this 
state is 
dynamically unstable and the molecules are 
rotating.\cite{Wasylishen:ssc85,Baikie:jmca:15}

Overall, we see that the nitrogen (most electronegative) side of the molecules 
is close to the iodines. The average distance of the 
N atom to nearest neighbor (n.n.) I atoms is $3.58\pm0.04\,\angstrom$/$ 
3.64\pm0.02\,\angstrom$ and of the C atom to n.n. I atoms is
$3.94\pm0.02\,\angstrom$/ 
$4.02\pm0.09\,\angstrom$, for the orthorhombic/tetragonal structure, 
respectively. 
The N-H 
bonds align to the N-I directions, thereby forming hydrogen 
bonds.\cite{Lee:cc15,Lee:sr16} All H-I bonds with a 
length $\le 3\,\angstrom$ have been indicated in Fig. \ref{fig1a} by the dashed 
(orange) lines. This orientation reduces the 
electrostatic energy between the negatively 
charged iodines and positively charged molecule. It dominates 
over the intermolecular dipole-dipole interaction as can be seen in the 
tetragonal 
structure (with its rotated octahedra), where it stabilizes the 
FE 
pattern.

\begin{table}[!t]
\caption{Overview of structures evaluated by molecular dynamics. Super cells of 
$n\times{}n\times{}n$ cubic unit cells, starting with an unpolarized-random (R) 
or polarized-aligned (A) MA ordering pattern, have been propagated for T 
picoseconds in steps of dt femtoseconds, with hydrogen mass $\rm M_{H}$ and at 
different 
temperatures.}
\label{table1}
\begin{ruledtabular}
\begin{tabular}{lcccc}
Size & T (ps) & dt (fs) & $\rm M_{H}$ (a.u.)&Temp (K) \\
\hline
 $2\times2\times2$ (R1)& 86 & 2.0 & 4 & 300 \\
 $2\times2\times2$ (R2)& 63 & 2.0 & 4 & 300 \\
 $2\times2\times2$ (R3)& 64 & 2.0 & 4 & 300 \\
 $2\times2\times2$ (A)& 84 & 2.0 & 4 & 300 \\
$4\times4\times4$ (R)& 21 & 2.0 &4 & 100  \\
$4\times4\times4$ (R)& 27 & 2.0 &4 & 150  \\
$4\times4\times4$ (R)& 42 & 2.0 &4 & 200  \\
$4\times4\times4$ (R)& 24 & 2.0 &4 & 250  \\
$4\times4\times4$ (R)& 23 & 2.0 &4 & 300  \\
$4\times4\times4$ (A)& 31 & 2.0 &4 & 300  \\
$4\times4\times4$ (R)& 31 & 2.0 &4 & 400\\
$6\times6\times6$ (R)& 8 & 3.0 &8 & 150 \\
$6\times6\times6$ (R)& 13 & 3.0 &8 & 300 
\end{tabular}
\end{ruledtabular}
\end{table}

\begin{figure}[!t]
\includegraphics[width=8.4cm]{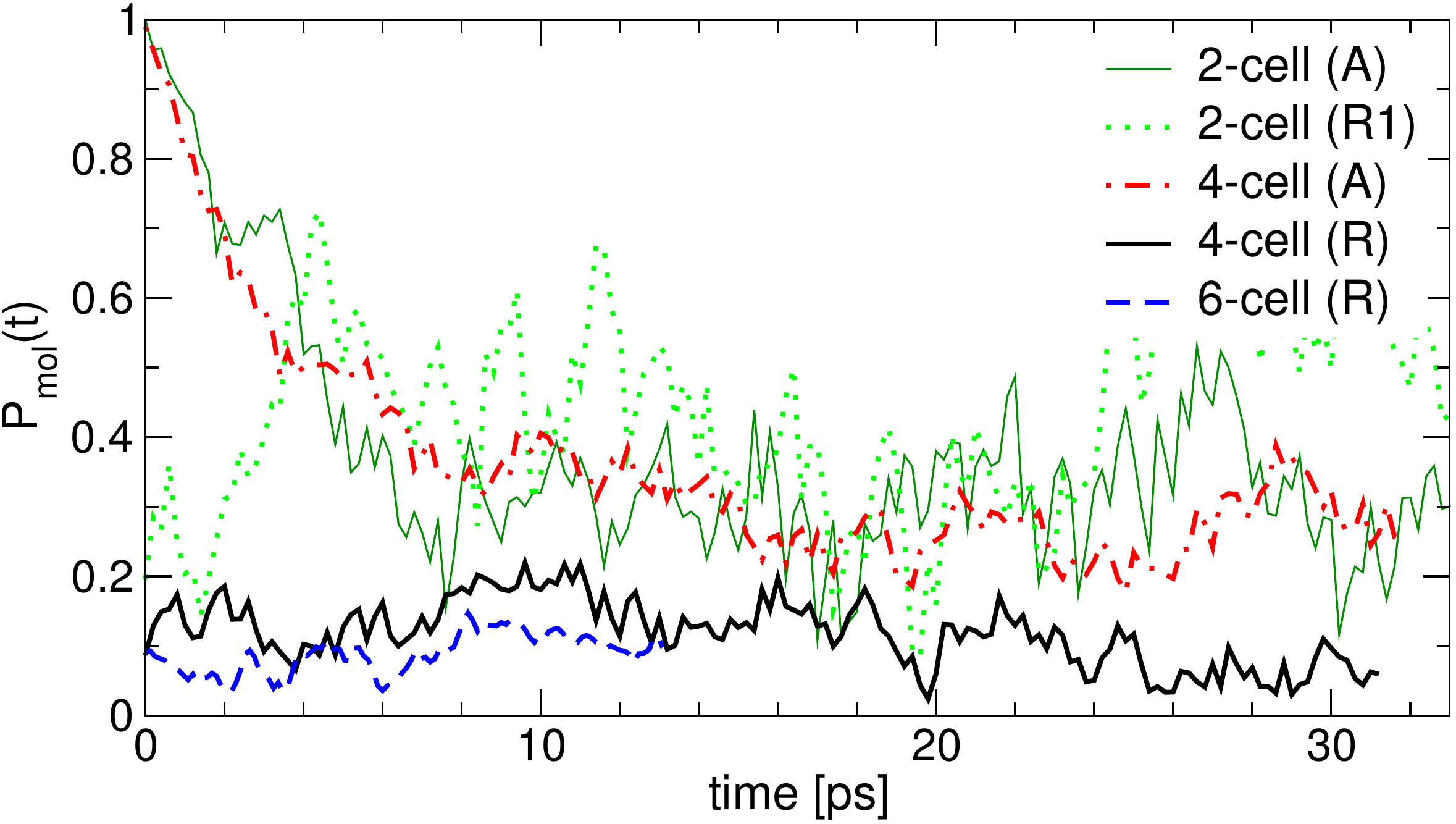}
\caption{(color online) Molecular polarization $\rm P_{\rm mol}(t)$ in the 
2-cell, 4-cell, and 6-cell at 300~K starting from an unpolarized-random (R) 
structure. 
The polarization resulting from a polarized-aligned (A) starting structure in 
the 
4-cell is given for comparison.}
\label{fig2}
\end{figure}

Now we move on to supercells and see if and how these ordering patterns 
prevail at finite temperature. Previous ab-initio MD calculations have used 
relatively small supercell structures of 8 formula 
units (cubic)\cite{Frost:aplm14,Quarti:pccp15} or 32 formula 
units (tetragonal)\cite{Quarti:pccp15,Quarti:ees15,Carignano:jpcc15}. Only 
Carignano \textit{et al.}\cite{Carignano:jpcc15} have systematically studied 
the effect of the supercell 
size and performed a calculation on a 108 formula unit (tetragonal) super 
cell. Using classical molecular dynamics Mattoni \textit{et al.} have studied a 
256 formula unit (orthorhombic) supercell.\cite{Mattoni:jpcc15} An overview of 
the supercells that we have evaluated at 
room temperature and below is presented in Table \ref{table1}. We have 
constructed $n\times{}n\times{}n$ supercells of the pseudo-cubic structure (1 
formula 
unit), thereby containing ${\rm N}=n^3$ molecules. This is hereafter referred 
to as the $n$-cell. The lattice parameters of the pseudo-cubic structure are 
fixed 
at the experimental values, while all 
internal coordinates have been fully relaxed at the DFT level. During the MD 
all atoms are allowed to move, but the cell volume and shape are kept fixed. 
Since we are evaluating large cells this approximation seems acceptable 
and the error diminishes going from $n=2$ to $n=6$. From the trajectory of 
all atoms we extract the orientation of all MA molecules (C-N bond vector) in 
polar 
coordinates projected on the unit sphere ($\phi_i(t),\theta_i(t)$). The 
coordinate frame is aligned to the 
Pb-I framework, as shown in Fig. \ref{fig1a}. Initially, all molecules in the 
super 
cells were randomly rotated over three axes to generate an unbiased 
starting structure. We have tested this in comparison to a 
starting 
structure with aligned molecules and found that this approach results in 
shorter 
equilibration times. For example, the first 20 ps of an MD 
trajectory at 300~K for an initially aligned 4-cell involve a reorientation 
of the molecules so that the initial long range order is practically 
destroyed. Starting from a random state avoids these long equilibration 
times. This can be seen in Figure \ref{fig2}, where the average 
polarization based on the molecular contribution only,
\begin{equation}
 {\rm P}_{\rm mol}(t)=\left|\frac{1}{\rm N}\sum_{i=1}^{\rm N} 
\mathbf{\hat{p}}_i(t)\right|,
\end{equation}
has been plotted for a polarized-aligned (A) and a unpolarized-random (R) 
starting structure. This is not the full polarization of the structure, 
which would include contributions from the cage, but 
merely the contribution of the intrinsic dipole moment vector 
($\mathbf{\hat{p}}_i$) of each molecule. Here we assume that the dipole of 
each molecule is equally strong and that it does not change in 
time. 
After initial equilibration time, both calculations show a 
fluctuating behavior 
around a mean value. The mean value depends on 
the starting configuration and does not converge to a 
common value within the available compute time. The higher degree of 
fluctuation might
indicate that the (A) system is not yet as well equilibrated as the (R) 
system.

\section{Dynamics of Individual Molecules}
\label{secB}

\begin{figure} [!h]
\includegraphics[width=8.4cm]{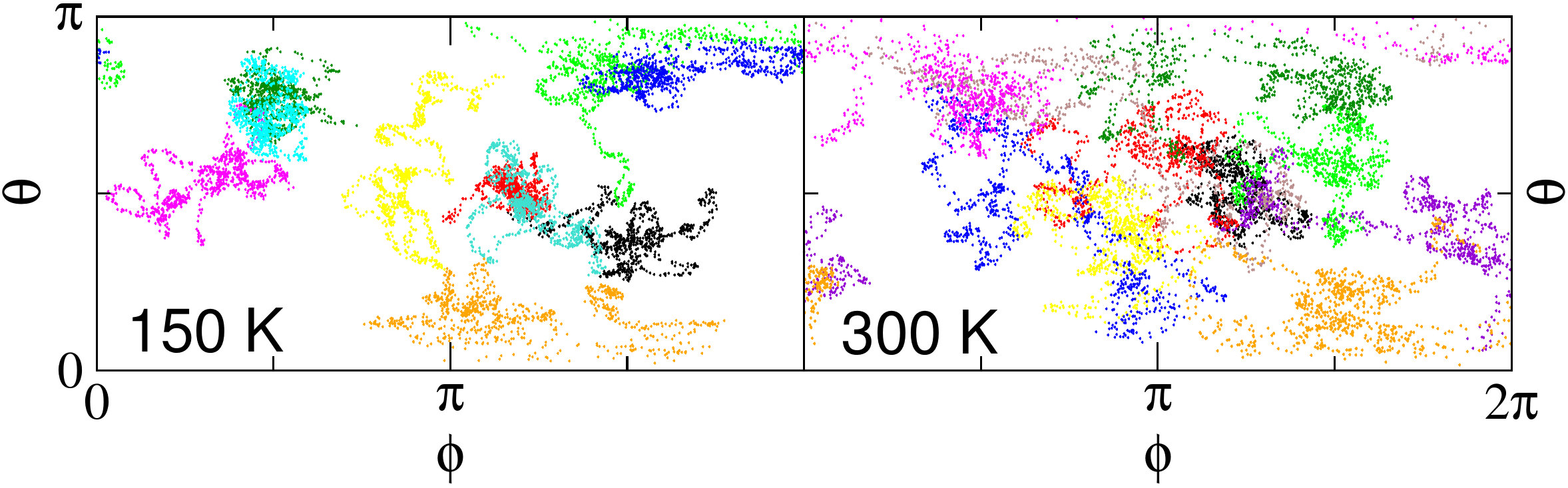}
\caption{(color online) The polar plot of 10 randomly chosen dipoles in 
the 6-cell shows the rotational mobility ($\phi_i(t),\theta_i(t)$) in a 
time frame of 8~ps at 150~K and 300~K.}
\label{fig3}
\end{figure}

We now want to determine how rotationally mobile the individual molecules are. 
In Figure 
\ref{fig3} the coordinates ($\phi_i(t),\theta_i(t)$) of 10 randomly 
chosen molecules in the 6-cell have been plotted over a time frame of 8~ps. 
The area covered by the paths increases with temperature. At 150~K the average 
orientation of most of the dipoles is maintained. At 300~K the covered area 
of the path ways is so large that reorientation of the molecules 
by $180^{\circ}$ is possible. How 
exactly the molecules move at 300~K is not yet clear, i.e., whether they 
perform free rotations or jump from one preferential orientation 
to the next, as was recently suggested.\cite{Leguy:natc15}

\begin{table}[!b]
\caption{The relaxation 
time ($\tau_{\rm mol}$) in picoseconds for a 
$n$-cell 
at 300~K staring from a polarized-aligned (A) and unpolarized-random (R) 
structures.}
\label{table2}
\begin{ruledtabular}
\begin{tabular}{l|ccccccc}
$n$ &  2 (A)&2 (R1)&2 (R2)&2 (R3)&4 (A)&4 (R)&6 (R) \\
\hline
$\tau_{\rm mol}$ & 4.8& 7.6&7.6&11.3&8.5&6.8&5.4
\end{tabular}
\end{ruledtabular}
\end{table}

\begin{figure} 
\includegraphics[width=8.2cm]{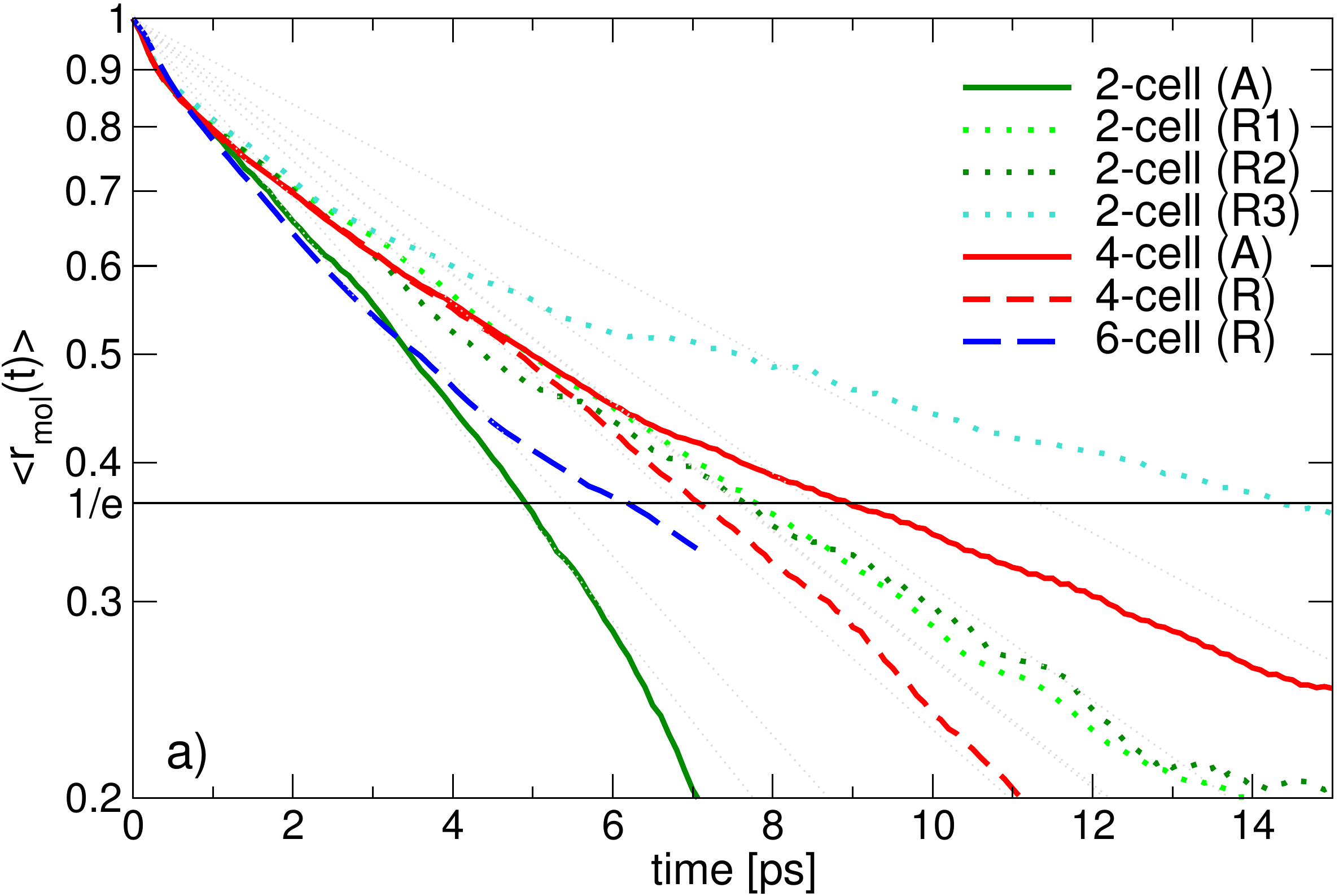}
\includegraphics[width=8.4cm]{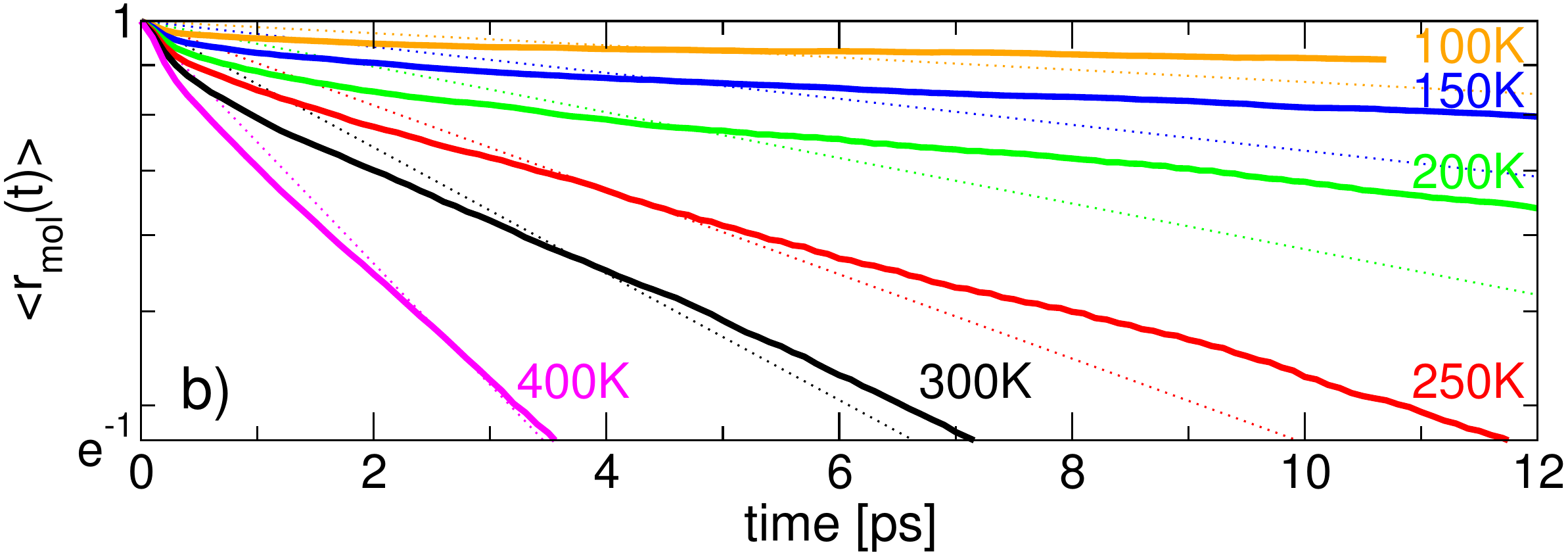}
\caption{(color online) Autocorrelation function $\left<{\rm r}_{\rm 
mol}(t)\right>$ of MA 
molecules in semi-log plot. a) At 300~K in the 2,4, and 6-cell structures 
 starting from random (R) and aligned (A) configuration of 
the molecules. b)  
At various temperatures in between 100-400~K in the 4-cell. The 
dashed lines indicate exponential fits.}
\label{fig4}
\end{figure}

The average reorientation time ($\tau_{\rm mol}$) of the $\rm N$ molecules 
can be extracted from the autocorrelation function
\begin{equation}
{\rm r}_{\rm mol}(t-t_0)=\frac{1}{\rm N}\sum_{i=1}^{\rm N} 
\mathbf{\hat{p}}_i(t_0)\cdot\mathbf{\hat{p}}_i(t),
\end{equation}
where the dot product traces the change of the molecule's orientation in time 
with respect to its orientation at $t=t_0$. The autocorrelation function 
is smoothed by applying a time averaging over $N_T$ different starting times 
$t_0$,
\begin{equation}
\left<{\rm r}_{\rm mol}(t)\right>=\frac{1}{N_T {\rm N}} \sum_{j=0}^{N_T-1}
\sum_{i=1}^{\rm N} 
\mathbf{\hat{p}}_i(\Delta t\,j)\cdot\mathbf{\hat{p}}_i(t+\Delta t\,j),
\label{eq3}
\end{equation}
where $\Delta t=\frac{\rm 
T}{2(N_T-1)}$ for $N_T>1$. Within this definition 
$\left<{\rm r}_{\rm mol}(t)\right>$ is computable only for half of the total MD 
trajectory time $\rm T$. In Figure \ref{fig4}~a), the 
autocorrelation functions of MA 
molecules in the 2,4, 
and 6-cell structure at 300~K are shown. $\tau_{\rm mol}$ is the 
time at which the autocorrelation function reaches the value 
$\nicefrac{1}{\rm e}$.  
Even though the autocorrelation functions are averaged with Eq.~\ref{eq3} and 
long MD trajectories were used, the relaxation times are different and depend 
on 
the starting configuration. In the 2-cell, using the aligned (A) molecules as a 
starting structure results in faster dynamics than randomly (R) oriented 
molecules. However, even between different random structures, R1, R2 and R3 the 
$\tau_{\rm mol}$ values differ. Some starting 
configurations are energetically more disfavored and might therefore be easier 
to break 
down. This means that with the present results for the 2-cell, we have an 
uncertainty in the $\tau_{\rm mol}$ value of about $\pm$3~ps. 
To reduce the uncertainty, one has to improve the statistics. This can be 
done by averaging over many more MD trajectories using different random 
starting configurations.

The values $\tau_{\rm mol}$ for the different systems have been tabulated in 
Table \ref{table2}. They have been obtained by fitting an exponentially 
decaying function to the curves.  These calculations 
show that $\tau_{\rm mol}$ in the 4-cell is at least of the same order as in 
the 2-cell, however with the present data we cannot conclude with certainty 
that the reorientation time is not 
affected by finite size effects. Note that the $\tau_{\rm mol}$ values are 
overestimated 
because the hydrogen masses have been raised. This raises the molecule's moment 
of 
inertia, while the DFT forces 
remain unchanged. The 6-cell shows a 
slightly lower reorientation time, which is possibly an artifact from the  
lower Pb and I masses (see Computational Details Section). 

Figure \ref{fig4}~b) shows 
the 
autocorrelation function in 
the 4-cell for various temperatures. The corresponding $\tau_{\rm mol}$ values 
are 3.5, 6.9, 9.9, 18, 32 and 69~ps for 400, 300, 250, 200, 150 and 100~K, 
respectively. For temperatures below 200~K these numbers do not have 
a significant meaning. At these temperatures the autocorrelation function 
is not well described by a simple exponential.\cite{Mattoni:jpcc15} Analysis 
with jump models on temperature dependent neutron scattering data shows 
relaxation time values of the same order.\cite{Chen:pccp15} Compared to this 
experiment here the presented results in the mid-temperature phase (150-300~K) 
are slightly lower.

\begin{figure}[!t]
\includegraphics[width=8.5cm]{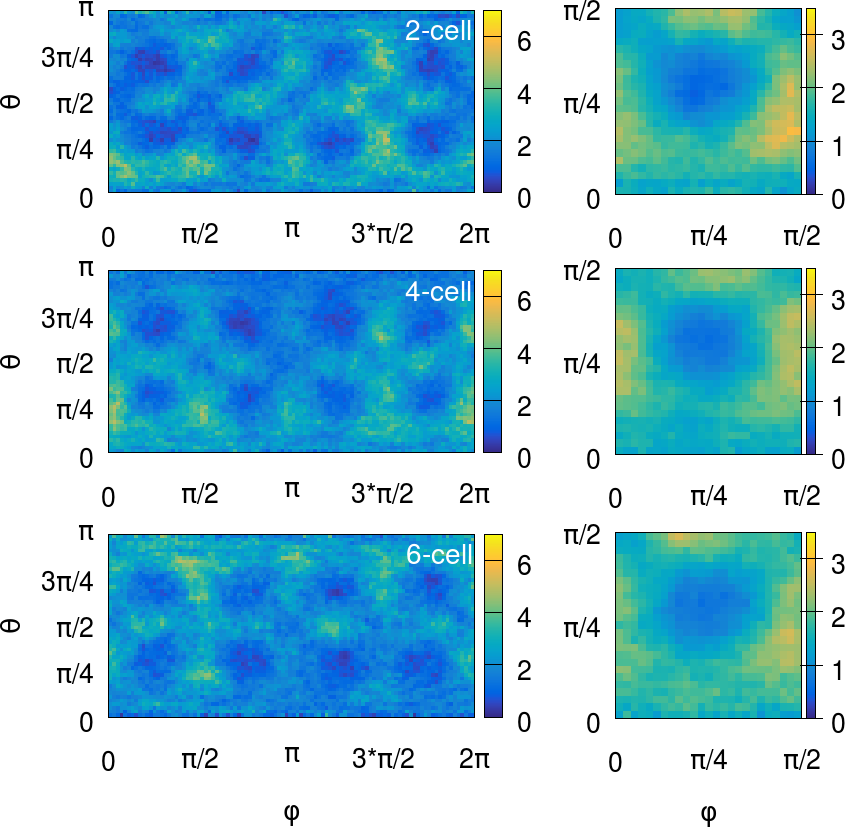}
\caption{(color online) Polar angle distribution of the MA molecules at 300~K 
in 
the 
2-cell (top), 4-cell (middle) and 6-cell (bottom). The plots show 
the full rotational space (left) and the down-folded plot (right) into 
$\nicefrac{1}{8}$ of the full space.}
\label{fig5}
\end{figure}

\begin{figure*} 
\includegraphics[width=17.5cm]{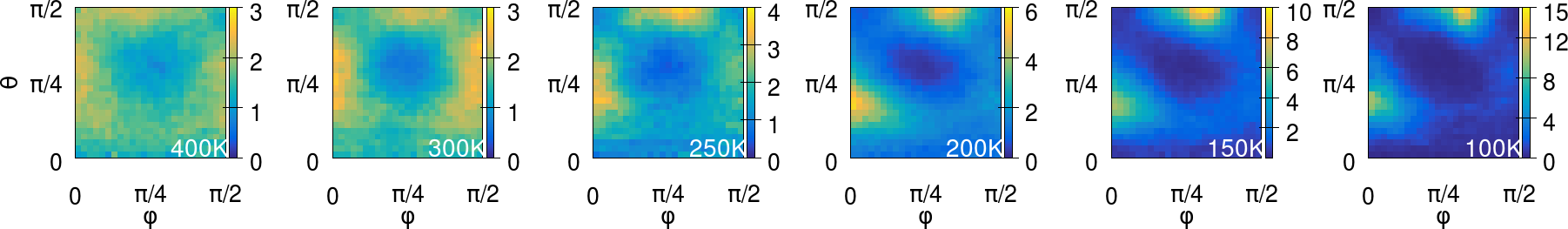}
\caption{(color online) From left to right, down-folded polar angle 
distribution of the MA 
molecules at 400, 300, 250, 200, 150, 100~K in the 4-cell.}
\label{fig6}
\end{figure*}

\begin{figure}[!t]
\includegraphics[width=7cm]{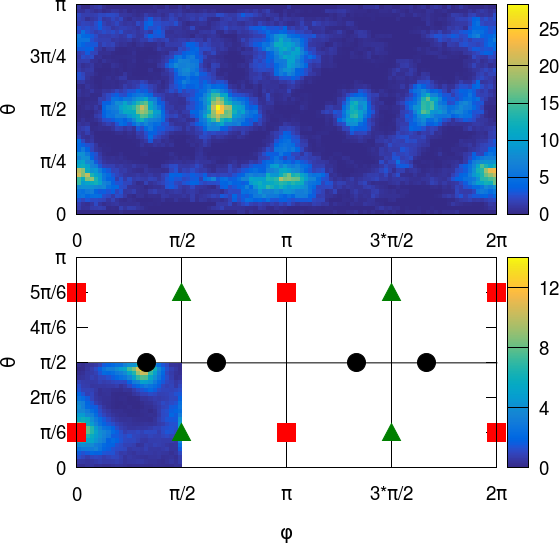}
\caption{(color online) \textit{Top:} Polar angle distribution of the MA 
molecules at 100~K in the 4-cell. \textit{Bottom:} Ideal molecular 
ordering of the orthorhombic structure with the molecules in the $xy$-plane 
(black circles), the $xz$-plane (red squares) and in the $yz$-plane (green 
triangles) with a $\nicefrac{\pi}{3}$ 
relative orientation. The down-folded 
polar angle distribution is included for 
comparison.}
\label{fig100Ksym}
\end{figure}

In order to understand the molecular order at 300~K, 
the polar distribution of the orientation ($\phi_i(t),\theta_i(t)$) of all 
molecules in the cell over the entire simulation time ($\rm T$) is plotted in 
Figure \ref{fig5}. 
The distribution is plotted for the 2,~4 and 6-cell.  
Provided that a sufficiently long simulation time is taken at 300~K, the polar 
distributions are independent of the precise starting configuration. Here,  MD 
trajectories from different starting molecular orderings (A/R) were combined 
in a single polar plot. The polar distributions of 
trajectories starting from the aligned and the three random structures for the 
2-cell have been combined (A+3R). For the 4-cell, the full trajectory starting 
from the random 
configuration and the second half of the trajectory of the aligned starting 
configuration have been combined (A+R). This removes the strong artificial bias 
of 
aligned ordering. The differences in 
the distribution between the 2-cell and the larger 4,6-cells  
are small. The remaining differences 
are predominantly related 
to the insufficient statistics. This means that the effective potential in 
which the molecules move does not depend strongly on cell size. (The effective 
potential 
can be obtained by a Boltzmann inversion of the polar distribution.) Combined 
with the foregoing observation of 
similar $\tau_{\rm mol}$ times this indicates that the dynamics of the 
individual molecules are roughly the same in the 2-, 4- and 6-cell. The 
polar plots presented in Fig. \ref{fig5} are in
agreement with the 300~K ab-initio/classical MD 
calculations in Refs. 
\onlinecite{Carignano:jpcc15}/\onlinecite{Mattoni:jpcc15}, respectively. The 
regular pattern in the polar plots shows the presence of cubic 
symmetry. Therefore we have down-folded the polar plots into a single octant, 
shown on the right. The hole in the middle of the down-folded plot corresponds 
to the room-diagonal orientation of 
the Pb-I cube. This orientation is systematically avoided by the MA molecules. 
Instead, the preferred direction is close to the face diagonal. The plot 
also shows that 
almost all orientations except the room-diagonal orientation have a fairly high 
probability. These results agree with the picture arising from NMR 
measurements. 
At room temperature the molecules are rotating freely in a potential 
with several preferred high symmetry directions.\cite{Baikie:jmca:15}

When the temperature of the system is lowered, a more 
anisotropic polar angle distribution appears. In Figure 
\ref{fig6} the down-folded polar plots of a 4-cell 
 at temperatures between 400 and 100~K are shown. When cooling 
down, the pattern changes from a ring with three spots, to just two spots below 
200~K. We are not able to pin-down the phase transition 
temperature to the orthorhombic/low-temperature phase (exp. 
$\sim$150~K\cite{Stoumpos:ic13}) 
exactly. Figure ~\ref{fig6} indicates that the low-temperature phase in our 
supercell approach is related to 
the freezing in of the molecules into two preferred orientations in the 
down-folded polar plot. The high 
temperature patterns show a similar polar distribution as the ones
calculated by the classical MD 
calculations of Mattoni \textit{et al.} in Ref.~\onlinecite{Mattoni:jpcc15}. 
However, for temperatures below 250 K there are some differences. The most 
prominent difference appears at 100~K, where we find two distinct spots in the 
down-folded polar plot instead of one. The first and most 
dominant spot is close to 
($\phi=\nicefrac{\pi}{3},\theta=\nicefrac{\pi}{2},$), a direction where the 
molecules are in the $xy$-plane and are slightly titled off the face 
diagonals 
of the Pb-I framework. The second spot close to 
($\phi=0,\theta=\nicefrac{\pi}{6}$) is absent in 
Ref.~\onlinecite{Mattoni:jpcc15}. What 
has 
happened is more easy to identify in the full polar plot, as shown in Figure 
\ref{fig100Ksym} (top). As mentioned before in the orthorhombic 
structure (Fig. \ref{fig1a}) all molecules lie flat 
in the $xy$-plane with a $\sim60^{\circ}$ angle to their neighbors and an AFE 
stacking in the $z$-direction. These four distinct orientations are presented 
by 
the black 
circles in the symmetrical polar plot of Fig. \ref{fig100Ksym} (bottom). The 
second spot also
describes such a pattern, but with the molecules in the $xz$-plane 
(presented by the red squares). This 
shows that, while cooling down the system, we acquired two competing 
orthorhombic phases ($xy$ and $xz$, circles and squares in Fig. 
\ref{fig100Ksym}) in the supercell. This low-temperature phase does not 
represent the ground state minimum, but as a result of the 
rapid cooling rate the system has been trapped in a local minimum. To avoid 
confusion we refer to this mixed phase in the supercell as the low-temperature 
phase. We 
furthermore note that the condensation into ordered phases with molecules lying 
in the $xy$ or $xz$ plane occurred even though we used the pseudo-cubic 
initial structure.

The general picture that appears in Fig. \ref{fig6} about the transition of the 
molecular order in the supercell from the low-temperature (100~K) to the 
high-temperature (300~K) phase is one of three competing orthorhombic phases, 
with molecules in the $xy$,~$xz$ and $yz$ planes. When the temperature 
increases, more than one phase can 
be present and the ideal spots start to blur out. This process goes on until a 
closed ring is formed for temperatures $>300$~K. Within the available 
simulation time, we have not observed a 'pure' tetragonal phase in the 
mid-temperature regime. According to the unit cell of Fig. 
\ref{fig1a} the 
tetragonal phase should result in four spots at 
$\{\phi=n\cdot\nicefrac{\pi}{2},\theta=\nicefrac{\pi}{3}\}$ with $n=0,1,2,3$. 
Therefore 
our mid-temperature phase is most likely to be a mixture between the 
orthorhombic phases rather than a representative of the tetragonal structure. 
We return to this point in Fig.~\ref{fig8}.

At this point we have 
presented the
orientation of the set of all molecules and their average reorientation times, 
but we do not know about the order between neighboring molecules. 
To this end, one needs to study the correlation between molecules.


\section{Correlation between molecules}
\label{secC}

In the previous two sections we have described the behavior by averaging over 
all molecules (Sec.~\ref{secA}) and the behavior of each molecule individually 
(Sec.~\ref{secB}). In this section the correlation between the orientation of 
a molecule and its nearest neighbors (n.n.) is studied both in time 
(\textit{dynamical correlation}) and on time average (\textit{static 
correlation}). 

\subsection{Static correlation}
\begin{figure}[!h]
\includegraphics[width=8.4cm]{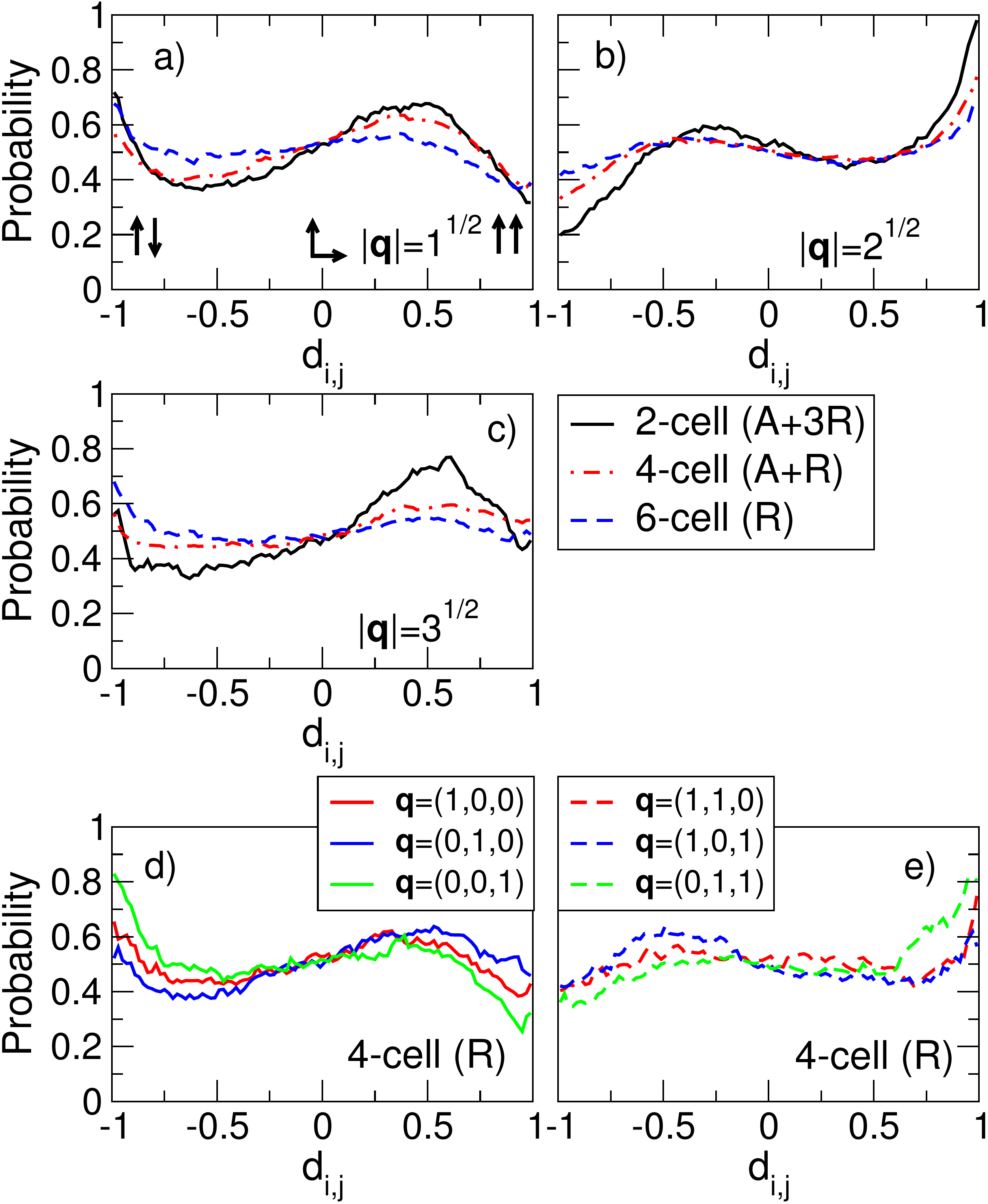}
\caption{(color online) Static correlation between molecules at 300~K in the 
a) 1st, b) 2nd and c) 3rd n.n. shells expressed by the distribution 
of ${\rm d}_{i,j}$, going from AFE ($\uparrow\downarrow$) to FE 
($\uparrow\uparrow$) alignment. The red/blue/green lines correspond to 
the 2-cell/4-cell/6-cell, respectively. The three different directions  
for the d) 1st (cubic axis) and the e) 2nd (face-diagonal) n.n. in the 4-cell 
at 300~K.  (Integral of the ${\rm d}_{i,j}$ distributions are 
normalised to one.)}
\label{fig7}
\end{figure}

The dot product between 
the unitary dipole vectors of molecules $i$ and $j$
\begin{equation}
 {\rm d}_{i,j}(t)=\mathbf{\hat{p}}_i(t)\cdot\mathbf{\hat{p}}_j(t),
 \end{equation}
indicates whether the molecules are arranged to be parallel ($1$), orthogonal 
($0$) or
anti-parallel ($-1$). All other relative orientations are within the 
range ($[-1,1]$). We 
focus on the orientation of one molecule with respect to molecules at 
different distances/directions: $|\mathbf{q}|=|(k,l,m)|$. Here $k,l,m$ are 
integers ($1,\ldots,n$ for an $n$-cell) describing the connecting vector 
between 
molecules $i$ and $j$ on a cubic grid. 
Figure \ref{fig7} shows 
the time averaged distribution of ${\rm d}_{i,j}$ over the whole trajectory of 
the MD 
simulation for the 2-, 4-, and 6-cell at 300~K. The first 
observation is 
that all relative orientations occur in all simulated $n$-cells. Second, 
 the shape of the distribution, characterized by the 
positions of the maxima and minima, does not depend very significantly on the 
cell size. Uncorrelated 
dipoles would result in a flat distribution, and therefore some correlation 
prevails 
at 300~K. The differences between the 2-, 4- and 6-cell  are 
related to the limited simulation time for the large supercells. Initially the 
molecules have a random orientation, and therefore it takes simulation time to 
bring the flat distribution to the converged attenuated distribution. For the 
computationally less expensive 2-cell calculations these results are very 
close to the fully converged distribution for $|\mathbf{q}|=1$ and 
$\sqrt{2}$. Note that a combined trajectory of $\sim 300$~ps corresponding to 
four MD runs starting from different molecular ordering (A+3R) was necessary. 

In the 1st n.n. shell ($|\mathbf{q}|=\sqrt{1}$) of Fig. \ref{fig7}, one 'sharp' 
peak at 
${\rm d}_{i,j}=-1$ is observed, here the molecules are AFE aligned. A 
second broad peak is found at ${\rm d}_{i,j}=0.4$, indicating a 
$\sim66^{\circ}$ angle between the neighboring dipoles. Going to the 2nd n.n. 
shell ($|\mathbf{q}|=\sqrt{2}$), a similar pattern is observed, but mirrored. 
The sharp peak at ${\rm d}_{i,j}=1$ indicates a 
FE alignment between a molecule and its 2nd n.n. The broad peak is also 
observed, but is now at ${\rm d}_{i,j}=-0.4$ indicating a $\sim114^{\circ}$ 
angle.  For the 4- and 6-cells, the third distribution 
($|\mathbf{q}|=\sqrt{3}$) is almost exactly the 
same 
as the first. When going even further to the 4th ($|\mathbf{q}|=\sqrt{4}$) and 
7th ($|\mathbf{q}|=\sqrt{8}$) shells the same pattern prevails. This 
alternating 
or 'mirror' pattern is not an effect of the supercell size, since it is does 
not diminish going from the 4- to the 6-cell. 

The distributions in Figs. 
\ref{fig7}~a-c) show an average over the 
different directions, all neighbors at the same distance have been added. 
This is possible because there is almost no cubic symmetry breaking, as 
can be seen in Figs. \ref{fig7}~d-e). The static correlation along the 
cubic axis ($\mathbf{q}_x,\mathbf{q}_y$ and $\mathbf{q}_z$) is shown in Fig. 
\ref{fig7}~d) and for all 
face-diagonal neighbors ($\mathbf{q}_{xy},\mathbf{q}_{yz}$ and 
$\mathbf{q}_{xz}$ separately) in Fig. \ref{fig7}~e). The curves 
have the same shapes and maxima, but the relative 
intensity per peak varies slightly.

\begin{figure}[!t]
\includegraphics[width=8.4cm]{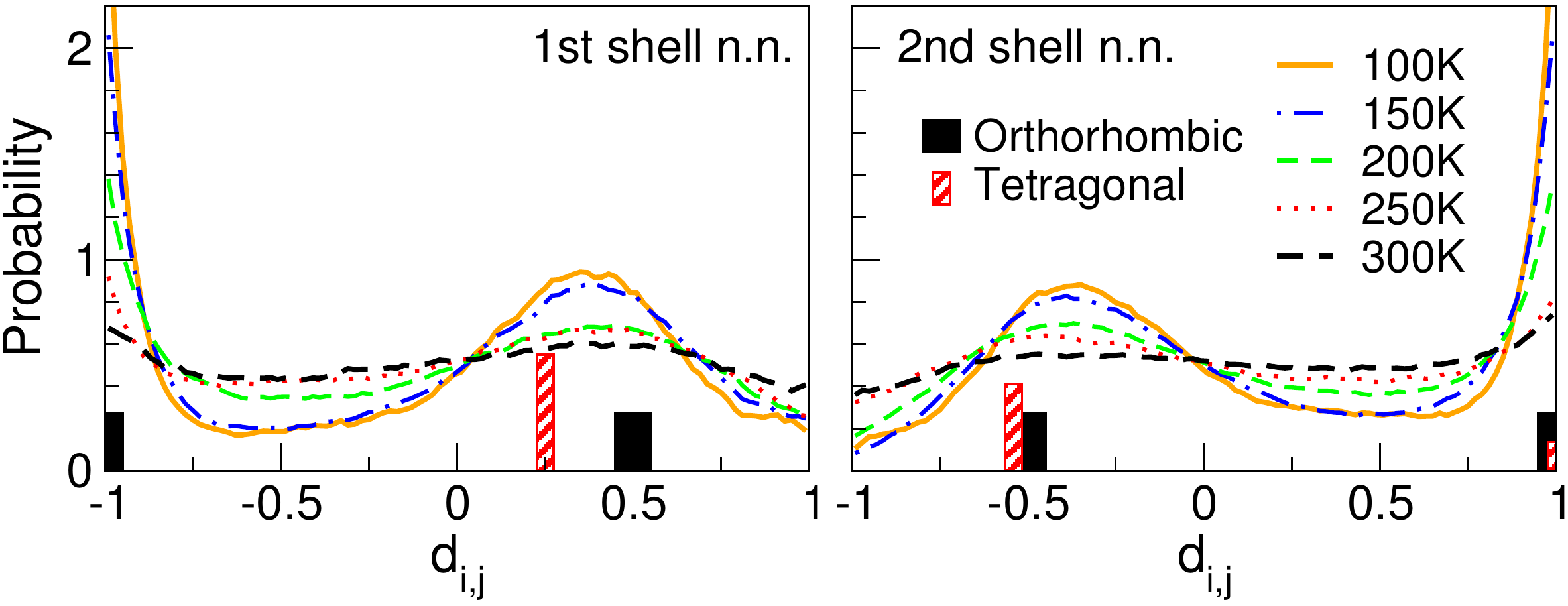}
\caption{(color online) Static correlation between molecules in the 1st (left) 
and 2nd 
(right) n.n. shell of the 4-cell at temperatures between 100 and 300~K 
expressed 
by 
the distribution of ${\rm d}_{i,j}$. The dot product values corresponding to 
the ideal molecular ordering pattern in the orthorhombic/tetragonal phase 
are shown (in arb. units) by the filled/striped histograms, respectively.}
\label{fig8}
\end{figure}

Analysis of Fig. \ref{fig7} confirms the picture deduced from the polar 
angle plots, i.e., that the molecular ordering pattern at 300~K is a 
mixture of orthorhombic phases smeared out by the high temperature. 
Plots of the distribution in the 4-cell at different temperatures further 
justify this view. The distributions for the 1st and 2nd n.n. shell are 
shown in 
Figure \ref{fig8}. For comparison the distribution 
in the perfect orthorhombic and tetragonal cells of Fig. \ref{fig1a} has been 
plotted by the solid (black) and dashed (red) histograms, respectively. When 
the system is cooled down the sharp AFE and broad 
$\sim60^{\circ}$ 
orientation peaks grow in relative intensity. These 
two peaks correspond to the 
molecular ordering pattern in the orthorhombic phase and are the same for the 
$xy$, $yz$ and $xz$-orientations. However, stress between the different 
phases in the supercell and the finite temperature can explain the shift and 
broadening of 
the peaks. The 
ratio between the most and 
least likely orientation is simultaneously increased from a factor of 2 at 
300~K 
to a factor of 25 at 100~K, indicating the freezing in of a fixed ordered 
pattern going hand in hand with an increase of static correlation. Again, no 
clear 
signature of the tetragonal phase is found in the 4-cell. 

\subsection{Dynamical correlation}
\begin{figure}[!t]
\includegraphics[width=8.4cm]{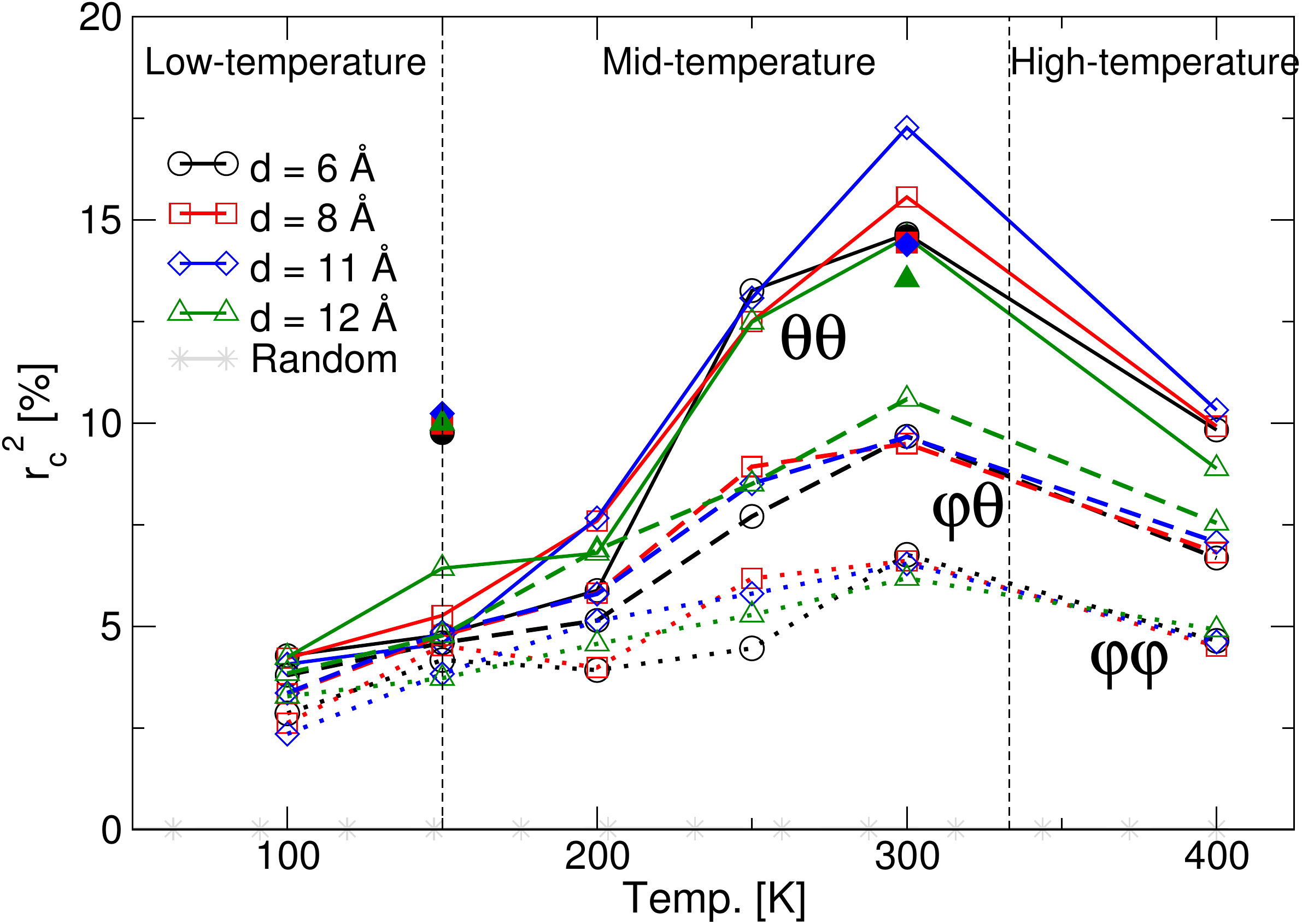}
\caption{(color online) Dynamical correlation between molecules in the first 
four n.n. shells of the 4-cell at temperatures between 100 and 300K expressed 
by square Pearson correlation 
coefficient (${\rm r_{c}^2}$) for $\theta_i\theta_j$, $\phi_i\theta_j$ and 
$\phi_i\phi_j$. The 
filled symbols correspond to the $\theta_i\theta_j$ 
correlation in the 6-cell. Experimental phase transition 
temperatures have been indicated by the dashed lines.}
\label{fig8b}
\end{figure}

As a measure of dynamical correlation between the MA molecules, we have 
calculated the Pearson correlation coefficient of the $\theta_i(t)$ 
trajectories of neighboring molecules. Here we use the square 
Pearson correlation 
coefficient
\begin{equation}
 {\rm r_{i,j}^2}=\left(\frac{\sum_{t=t_0}^{\rm T} 
(\theta_i(t)-\bar{\theta}_i)(\theta_j(t)-\bar{\theta}_j)}{\sigma_i^2\sigma_j^2}
\right)^2,
\end{equation}
with $\sigma^2$ the variance and $\bar{\theta}$ the mean of 
$\theta(t)$. It is an estimate of the fraction of the two signals showing 
linear correlation. In Figure \ref{fig8b} the average values ${\rm r_{c}^2}$ 
are 
plotted. This is the average ${{\rm r}_{i,j}^2}$ over all pairs $i,j$ with 
distance $d=|\mathbf{r}_i-\mathbf{r}_j|$ and ranges between 0-100\%. 12~ps long 
trajectories of the 4-cell MD calculations at different temperatures have been 
used. Qualitatively the same picture appears when the dynamical correlation of 
$\phi_i(t)$ with $\phi_j(t)$ (dashed lines) or $\phi_i(t)$ with $\theta_j(t)$ 
(dotted lines) are calculated. Random dipoles 
placed on the 4-cell grid result in a mere $\sim$0.02\% correlation. 
Remarkably, 
the dynamical correlation is temperature dependent and, even 
more remarkably, is independent of the distance, at least in 
the 4-cell considered here. This means that the movement 
of the molecules in the mid-temperature range is partly a collective motion. A 
recent Monte Carlo simulation based on 
a simplified model Hamiltonian of interacting MA molecules in MAPbBr$_3$ 
suggested correlation in clusters of 3 neighboring molecules.\cite{Motta:prb16} 
We cannot predict an effective rotation cluster size (because of the limited 
supercell size), but in the 4-cell it seems to be greater than 3. The 
filled symbols in Fig. \ref{fig8b} correspond to the $\theta_i\theta_j$ 
correlation in the 6-cell. At 300~K the same level of dynamic correlation as in 
the 4-cell is found, but at 150~K the value is larger. Presumably this is 
caused by the very limited simulation time and therefore too short 
equilibration time.

\section{Discussion}
\label{secE}
In this next to last section, we would like to hypothesize about the 
origins of 
the dynamics of the MA molecules. J. Li $et.al.$ already referred to it in Ref. 
\onlinecite{Jingrui:prb16} as the 'chicken and egg paradox' present 
in this system. Do the molecules determine the rotations of the PbI$_6$ 
octahedra or do the molecules follow the octahedra? Their conclusion was that 
the two mechanisms work synergetically. The results presented 
here shed new light on this issue from a time and temperature dependent 
viewpoint. We focus now on the, at first sight, peculiar temperature 
dependence of the dynamical correlation in Fig.~\ref{fig8b}.  
A peak in the dynamical correlation at finite temperature is unusual. In most 
cases dynamical correlation decreases when temperature goes up. For example, 
the correlation between two next nearest neighboring atoms in a cubic crystal 
performing the (very correlated) motion associated to a phonon mode. If the 
temperature increases, modes tend to broaden and higher lying modes become 
occupied. This will lower the dynamical correlation between the atoms. 
However in Fig.~\ref{fig8b}, in the 
low-temperature phase, the 
molecules do not posses enough kinetic energy to overcome the potential 
barrier imposed by the Pb-I framework. Therefore at low 
temperatures, where the molecules only oscillate (random thermal motion) around 
their equilibrium positions and rotate along the C-N axis, the motion between 
neighboring molecules is less 
correlated. In the intermediate temperature phase the reorientation times 
($\tau_{\rm mol}$) become 
finite. The molecules then have enough kinetic energy to 
overcome the potential barrier. However, because neighboring octahedra are 
rotated with respect to each other in the low and mid-temperature range, the 
potential landscape in these phases is expected to be less isotropic compared 
to the high-temperature (pseudo-cubic) phase. It is therefore likely 
that a molecule in the mid-temperature regime can only rotate if (some of) its 
surrounding molecules rotate 
as well. The collective reorientation prevents large stress and strain 
interactions with the Pb-I framework. When the temperature is raised (but 
remaining in the mid-temperature regime) more (concerted) reorientation 
pathways 
become available and the average 
level of dynamical correlation increases. When the temperature is raised even 
further into the high-temperature phase, the kinetic energy lies above the 
barriers imposed by the now more 
isotropic potential landscape of the cage and the molecules rotate 
independently. Overall, we suggest that the Pb-I framework is 
the mediator here, coupling molecule $i$ with $j$ and not the electrostatic 
dipole-dipole interaction, which would be distance dependent.

\section{Conclusion}
\label{secF}
Large supercell ab-initio molecular dynamics calculations of the perovskite 
solar cell material MAPbI$_3$ have been performed. The dynamics of the 
 initially randomly oriented MA 
molecules has been studied and carefully analyzed with respect to the supercell 
size. 
 We find that at 300~K the $2\times2\times2$ (2-cell, 96 
atoms), 
$4\times4\times4$ supercell (4-cell, 768 atoms) as well as the 
$6\times6\times6$ (6-cell, 2592 atoms) supercells result in very similar polar 
angle distributions of the molecules if the 
molecular dynamics trajectories are long enough.
However, a small supercell (2-cell) can artificially constrain the dynamics 
of the molecules,  it is more sensitive to the initial 
random orientations of the molecules. Therefore either long ($>100$~ps) 
trajectories have to be calculated or a combination of multiple shorter 
runs starting from 
different independent random starting configurations has to be made. The 
average reorientation times ($\tau_{\rm mol}$) in the three supercells are of 
the same 
order. The molecules can rotate with a $\tau_{\rm mol}$ time of 
$\sim$~7~ps. This is an upper bound because increased hydrogen masses have been 
used. The molecules systematically avoid the room-diagonal 
orientation and prefer the (slightly tilted off) face-diagonal 
orientations of the 'cubical' Pb-I cage.

At room temperature all relative 
orientations between neighboring molecules can occur. However, signs of an 
ordering pattern related to the orthorhombic structure 
smeared out by the high temperature are visible. When the temperature is 
lowered to 
100~K the molecular ordering pattern condenses into a mixture of orthorhombic 
phases within the supercell. The dynamical correlation, expressed by the 
Pearson correlation coefficient between the $\theta(t)$ trajectories of 
neighboring molecules in the 4-cell, shows, in the considered temperature range 
of 100-400~K, a value between 5-15\%. A purely random system of the same size 
would result in a mere 0.02\% correlation. At room temperature a maximum in the 
dynamical correlation is observed. In this mid-temperature phase, the 
molecules possess enough 
thermal energy to rotate and this process is slow enough to couple to neighbors 
via the Pb-I cage. The range of the coupling extends throughout the 
entire considered supercell. This results in (partially) collective motion of 
neighboring molecules driven by stress and strain interactions which are 
mediated by the Pb-I framework. The role of intermolecular dipole-dipole 
interactions is most likely negligible in this process. At lower 
temperatures, the motions are less correlated, because the molecules randomly 
oscillate around  their fixed orientations. Also at higher 
temperatures there is less correlation, because the molecules spin around 
freely.

\acknowledgements{
MB, CF, AK and DDS acknowledge funding by the joint project of the Indian 
Department of Science and Technology (DST) and the Austrian Science Fund (FWF): 
I1490-N19 (INDOX). GK and CF acknowledge funding by Austrian Science Fund (FWF): 
F4102-N28, F4115-N28 (SFB ViCoM). The calculations were performed at the Vienna 
Scientific Cluster (VSC-3). VESTA\cite{VESTA} was used for Fig.~\ref{fig1a}.
}

\end{document}